\documentstyle[twoside,epsf]{article}

\input ibvs2.sty

\def\etal{{\it et al.}}
\newbox\grsign \setbox\grsign=\hbox{$>$}
\newdimen\grdimen \grdimen=\ht\grsign
\newbox\laxbox \newbox\gaxbox
\setbox\gaxbox=\hbox{\raise.5ex\hbox{$>$}\llap
     {\lower.5ex\hbox{$\sim$}}}\ht1=\grdimen\dp1=0pt
\setbox\laxbox=\hbox{\raise.5ex\hbox{$<$}\llap
     {\lower.5ex\hbox{$\sim$}}}\ht2=\grdimen\dp2=0pt

\def\lax{\mathrel{\copy\laxbox}}

\begin{document}
\IBVShead{4622}{?? August 1998}

\IBVStitle{S 10943 Vulpeculae: A New ROSAT Selected Dwarf Nova,}
\IBVStitle{probably of SU Ursae Majoris Subclass}

\IBVSauth{G.A. Richter$^1$, J. Greiner$^2$, P. Kroll$^1$}
\IBVSinst{Sternwarte Sonneberg, 96515 Sonneberg, Germany, 
  e-mail: Peter.Kroll@STW.TU-Ilmenau.DE}
\IBVSinst{Astrophysical Institute Potsdam, An der Sternwarte 16, 
   14482 Potsdam, Germany, e-mail: jgreiner@aip.de}

\IBVSobj{S 10943 Vul, RX J1953.1+2115, TU Men}

\begintext

As part of our program of investigating the optical long-term behaviour
of selected ROSAT X-ray sources  we studied the X-ray source 
RX\,J1953.1+2115 in more detail. It was discovered during a 360 sec. scanning
observation on Oct. 18--20, 1990 during the ROSAT all-sky survey at a mean 
count rate of 0.024$\pm$0.006 cts/sec. The hardness ratios HR1 = 
(N$_{\rm 52-201}$ -- N$_{\rm 11-41}$)/(N$_{\rm 11-41}$ + N$_{\rm 52-201}$) = 
1.00$\pm$0.30
(where N$_{\rm a-b}$ denotes the number of counts in ROSAT's position 
sensitive proportional counter between
channel a  and channel b) and 
HR2 = (N$_{\rm 91-200}$ -- N$_{\rm 50-90}$)/N$_{\rm 50-200}$ = --0.17$\pm$0.38,
though admittedly purely constrained due to the low number of counts,
suggest a moderately hard, but absorbed spectrum. Assuming a thermal
bremsstrahlung spectrum with kT = 2 keV, the unabsorbed flux in the
0.1--2.4 keV band ranges between 2$\times$10$^{-13}$ erg/cm$^2$/s
(for an assumed absorbing column of 
$N_{\rm H}$=1$\times$10$^{20}$ cm$^{-2}$) up to 
1.5$\times$10$^{-12}$ erg/cm$^2$/s (for the maximum galactic column 
in this direction of
$N_{\rm H}$=4.9$\times$10$^{21}$ cm$^{-2}$).

The best-fit X-ray position of RX\,J1953.1+2115 was determined as
RA = 19\hr53\mm05\fsec4, Decl. = +21\deg14\arcm31\arcs\ (equinox J2000.0)
with an error radius of 30\arcs. 
The Palomar Observatory Sky Survey prints revealed 17 objects within
this X-ray error radius which
were tested for variability on 250 archival plates of the Sonneberg
400\,mm astrograph (limiting magnitude $\sim$ 18\mm) and on 190 plates
of the 170\,mm triplet cameras (limiting magnitude $\sim$ 16\fmm5). All
these 17 objects proved to be constant (or always invisible on archival
plates) within the error of photometry with the exception of one 
($\approx$ 19\mm) slightly blue object heavily blended by
a 16\mm\ object only 6\arcs\ to the East (see Fig. 1). Due to the clear
variability exhibited by this object, it is assigned the number S\,10943
in the series of variable stars detected at Sonneberg Observatory.

S\,10943 shows outbursts up to 15\mm\ with a rather stable recurrence
time of 83.6 days. Table 1 gives a comprehensive summary of all the
outbursts found on the Sonneberg photographic plates.  
Shorter (less than 60 days) and longer (about 100
days) intervals are occasionally found, but are rare. Because of the
sporadic distribution of observations and the blending already
mentioned, the duration of the outbursts is difficult to estimate, but
both short ($<$ 10 days) and long ($>20$ days) outbursts seem to occur. 
A plate from Sep. 28, 1967 shows the object at a probable rise to  a 
superoutburst the evolution of which could be followed on 7 plates. 
On Oct. 28, i.e. 30 days later, the minimum brightness was not yet reached.
A series of 24 exposures, taken between 1995 May 1 and May 4,
cover the early decline (15\mm) of a long duration outburst, which
was not yet complete on May 21 (still about 1 mag above minimum
light). During those 4 days periodic brightness fluctuations
of small amplitude (0.2 mag) are  superimposed on a steady brightness
decrease of about 0.1 mag per day which may be interpreted as superhumps of
an SU Ursae Majoris star. The period length was determined to be P=0\fday1196
$\sim$ 2\fhr871. Two alternative values are also possible, but with smaller
probability: P=0\fday136 and P=0\fday107. With the superhump periods being as a
rule about 2-3\% longer than the orbital periods, we may expect an
orbital period near 2\fhr8 which is just at the upper border of the well-known
period-gap of cataclysmic variables.

\vskip 0.7cm
\centerline{Table 1. Observed eruptions (r = rise, m = maximum, d = decline)}
\vskip 3mm
\begin{center}
\begin{tabular}{|rlc|rlc|rlc|}
\hline
J.D. & m$_{\rm pg}$ & & J.D. & m$_{\rm pg}$ & & J.D. & m$_{\rm pg}$ & \\
\hline
2427710.318 & 16.5 & d ? & 2438378.251 & 16.0: & d & 2448888.395 & 15.9 & \\
9102.537 & 16.5: & r 	& 9003.400 & 16.2 &	& 9163.470 & 15.8 &	\\
9107.428 & 15.6: & d 	& 9347.405 & 17.0: & d	& 9504.464 & 15.6 & r ?	\\
9541.313 & 16.1 & 	& 9349.405 & 17.0: & d	& 9511.554 & 17.1: & d \\
9777.455 & 16.8 &	& 9762.358 & 16.6: & r	& 9839.521 & 14.9: & d	\\
9843.411 & 16.5: &	& 9765.310 & 16.0 & d	& 9839.535 & 15.8: & d	\\
2430442.616 & 15.0: & m & 9765.381 & 15.9 & d	& 9839.550 & 15.3 & d	\\
0614.354 & 15.0: & m	& 9767.297 & 16.0 & d	& 9840.505 & 15.6 & d	\\
0848.512 & 16? &	& 9789.261 & 17.7: & d	& 9840.523 & 15.4 & d	\\
1020.311 & 16? &	& 9789.304 & 17.3 & d	& 9840.542 & 15.3 & d	\\
1296.418 & 16.2 & r	& 9792.287 & 17.3: & d	& 9840.560 & 15.4 & d 	\\
1296.455 & 16.0 & r	& 2441917.367 & 15.7 & d & 9840.577 & 15.3 & d 	\\
1297.417 & 15.8 & m	& 1917.430 & 15.8 & d	& 9841.471 & 15.4 & d	\\
3160.431 & 15.6? &	& 2369.235 & 16.0 & 	& 9841.493 & 15.3 & d	\\
6073.424 & 16.4 &	& 4132.340 & 16.9 &	& 9841.508 & 15.4 & d	\\
6672.573 & 15.8 &	& 4132.359 & 16.8 &	& 9841.523 & 15.4 & d	\\
6815.373 & 16.8 &	& 6683.403 & 16.8 & r ? & 9841.537 & 15.4: & d	\\
7193.355 & 16.6: &	& 6699.339 & 15.6 & d	& 9841.552 & 15.4 & d 	\\
7576.444 & 16.6: &	& 6707.392 & 16.7 & d	& 9841.567 & 15.6 & d	\\
8268.386 & 18: & r	& 6708.390 & 16.4 & d	& 9841.581 & 15.5 & d 	\\
8282.326 & 17.9: & d	& 7365.493 & 16.3 & d	& 9842.467 & 15.4: & d	\\
8282.368 & 17.9: & d	& 7379.417 & 17.3: & d	& 9842.482 & 15.7 & d	\\
8283.327 & 17: & d	& 7381.428 & 17.1 & d	& 9842.497 & 15.5 & d	\\
8283.369 & 17.8: & d	& 7411.376 & 15.9 & m	& 9842.511 & 15.7 & d	\\
8284.364 & 18: & d 	& 7717.466 & 17.1: &	& 9842.526 & 15.7 & d	\\
8367.229 & 17: & r	& 7822.323 & 15.7 & d	& 9842.540 & 15.7 & d	\\
8370.241 & 15.9: & d	& 7823.260 & 15.9 & d	& 9842.555 & 15.7 & d	\\
8371.224 & 16.3 & d 	& 8096.448 & 16.8: & d	& 9842.569 & 15.4 & d	\\
8371.266 & 16.1: & d	& 8097.506 & 17.3: & d	& 9859.485 & 17.3 & d	\\
8372.257 & 16.1: & d	& 8804.472 & 17.0 &	& & & \\
\hline
\end{tabular}
\end{center}

TU Mensae (P$_{\rm SH}$=0\fday1262; Ritter \& Kolb 1998) is the only other
SU Ursae Majoris star with such a long period. The absolute magnitude of
TU Men during minimum brightness is $M_{\rm V}$=8.8 (Warner (1987).
Assuming a similar absolute magnitude for S\,10943 and using the
apparent brightness during minimum of $m_{\rm V} = m_{\rm B}$ = 19\fmm0
we derive an apparent distance modulus of 10.2 mag.
S\,10943 is situated at the border between 
area 1 (R.A. = 19\hr.6...19\hr9, Decl. = +15\deg...+25\deg) and 
area 3 (R.A. = $>$19\hr9, Decl. = +15\deg...+24\deg) for which Richter 
(1968) estimated the mean value of interstellar dust extinction to be 2.0 mag 
and 1.4 mag, respectively. The corresponding distance is 440 pc and 600 pc,
respectively. Alternatively, we can use the relation between orbital 
inclination and absolute magnitude for a comparison of the absolute magnitude 
of TU Men and S\,10943. With i = 65\deg\ for TU Men, and assuming
as an extreme (conservative) case i = 0\deg\ for S\,10943 (the lack of
eclipses on our plates suggests i $\lax$ 70\deg), the
difference of the absolute magnitude is
 $$\bigtriangleup M_{\rm V}(i) = -2.5 * {\rm log} (1 + 3/2 * {\rm cos}i) * 
    {\rm cos}i $$
(see e.g. formula 4 in Warner 1987, or also Paczynski \& 
Schwarzenberg-Czerny 1980). This results in a distance modulus of 11.0 mag,
or 580 pc and 700 pc, respectively. We therefore conclude that the most
likely distance of S\,10943 will be in the range of 400--700 pc.
At this distance the implied X-ray flux of 
6.2$\times$10$^{30}$--5$\times$10$^{31}$ (D/500 pc)$^2$ erg/s 
is well within the range of other SU UMa systems (van Teeseling \etal\ 1996).

\IBVSfig{14.3cm}{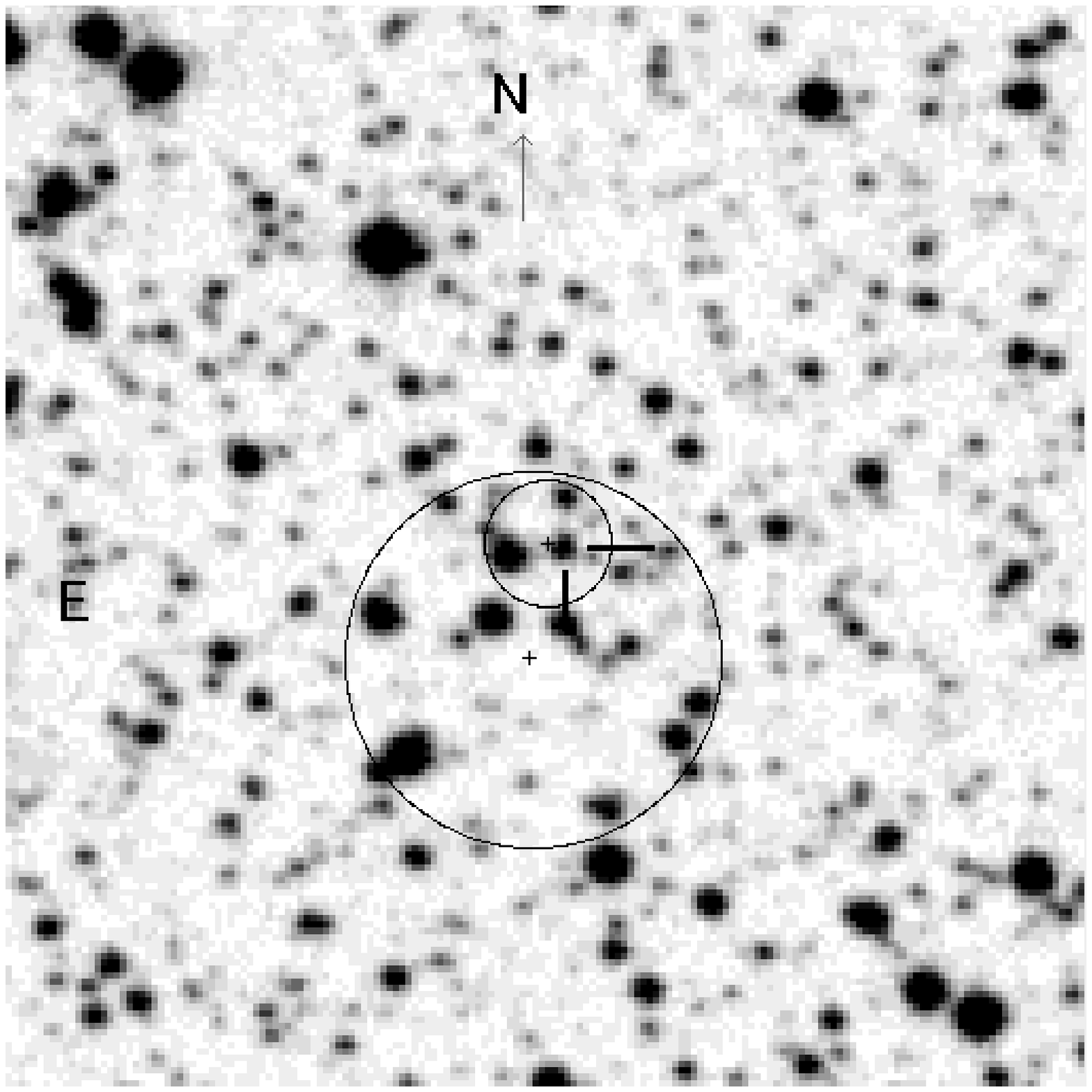}{A 3\arcm\ by 3\arcm\ part of the digitized
 sky-survey image (based on the red passband plate SF04200 taken on 
 9 Sep 1991) with the X-ray error circles of the ROSAT all-sky survey position 
 (large circle; 30\arcs\ radius) and the  HRI pointed observation 
 (small circle; 10\arcs) overplotted. S\,10943 Vul is marked by two 
 heavy dashes.}

The association of RX\,J1953.1+2115 with S\,10943 Vul has been
strengthened by the results of a recent ROSAT HRI observation. 
In the 6310 sec exposure on April 22--25, 1998 RX\,J1953.1+2115 was 
detected at a count rate of 0.0068$\pm$0.001 cts/s which is consistent
with the count rate during the ROSAT all-sky survey in 1990 given the 
factor 3 lower sensitivity of the HRI
as compared to the PSPC for X-ray sources with hard X-ray spectra.
This detection allowed an improved determination of the X-ray position
of RA = 19\hr53\mm05\fsec2, Decl. = +21\deg14\arcm50\arcs\ (equinox J2000,
$\pm$10\arcs).
The coordinate of S\,10943 Vul as measured on the Palomar blue print is
RA = 19\hr53\mm05\fsec0, Decl. = +21\deg14\arcm49\arcs\ (equinox J2000, 
$\pm$1\arcs), and thus is within 3\arcs\ of the X-ray position.
Fig. 1 shows the position of S\,10943 Vul relative to the two X-ray positions.

Thus, if the orbital period should be confirmed, S\,10943 would then be, 
together with TU Men, the SU Ursae Majoris star with the second largest 
superhump (and orbital) period known. 

\vskip 0.3cm
\normalsize
{\it Acknowledgements:}
This work was supported by the German Bundesministerium f\"ur Bildung,
Wissenschaft, Forschung und Technik under contract Nos. 05~2S0524 (GAR
and PK) as well as 50\,OR\,9201 and 50\,QQ\,9602 (JG).
Fig. 1 is based on photographic 
data of the National Geographic Society -- Palomar
Observatory Sky Survey (NGS-POSS) obtained using the Oschin Telescope on
Palomar Mountain.  
The Digitized Sky Survey was produced at the Space
Telescope Science Institute under US Government grant NAG W-2166.

\large
\references
Paczynski B., Schwarzenberg-Czerny A., 1980, {\it Acta Astron.} {\bf 30}, 127

Richter G.A., 1968, {\it Ver\"off. Sternwarte Sonneberg} {\bf 7}, 1

Ritter H., Kolb U., 1998, {\it A\&A Suppl}. {\bf 129}, 83

van Teeseling A., Beuermann K., Verbunt F., 1996, {\it A\&A} {\bf 315}, 467

Warner B., 1987, {\it MNRAS} {\bf 227}, 23

\end{document}